\documentclass[journal]{IEEEtran}

\usepackage{epsfig,amsmath,amssymb,epsf,amsthm,scalefnt, subfig}
\usepackage[table,dvipsnames]{xcolor}
\definecolor{myblue}{RGB}{0,51,160}
\definecolor{mypurple}{RGB}{142, 15, 237}
\definecolor{mygreen}{RGB}{75,148,112}
\usepackage{makecell}
\usepackage{float}
\usepackage{cite}
\usepackage{psfrag}
\usepackage{tabularx}   
\usepackage{array}      
\usepackage{siunitx}    
\usepackage{booktabs}
\usepackage{threeparttable}
\usepackage{arydshln}  
\usepackage{multirow}
\usepackage{makecell}
\usepackage{adjustbox}   
\usepackage{pifont}      
\usepackage[utf8]{inputenc}
\usepackage{enumitem}

\usepackage[normalem]{ulem} % \sout으로 주석 처리하는용

\newcolumntype{L}[1]{>{\raggedright\arraybackslash}p{#1}}
\newcolumntype{C}[1]{>{\centering\arraybackslash}p{#1}}
\newcolumntype{Y}{>{\raggedright\arraybackslash}X}

\newtheorem*{lemma*}{Lemma}

\def\b0{{\pmb{0}}} 

\theoremstyle{remark}

\renewenvironment{IEEEbiography}[1]
  {\IEEEbiographynophoto{#1}}
  {\endIEEEbiographynophoto}

\begin{document}

\title{Challenge-Response Authentication for LEO Satellite Channels: Exploiting Orbit-Specific Uniqueness}

\author{Jinyoung Lee, Stefano Tomasin, and Dong-Hyun Jung\\
\thanks{Jinyoung Lee is with the Division of Electronics and Electrical Information Engineering, National Korea Maritime \& Ocean University, Busan 49112, South Korea; Stefano Tomasin is with the Department of Information Engineering, University of Padova, Padova 35131, Italy; Dong-Hyun Jung (corresponding author) is with the School of Electronic Engineering, Soongsil University, Seoul 06978, South Korea.}
}

\maketitle
\begin{abstract}
The number of low Earth orbit (LEO) satellite constellations has grown rapidly in recent years, bringing a major change to global wireless communications. As LEO satellite links take on a growing role in critical services such as emergency communications, navigation, wide-area data collection, and military operations, keeping these links secure has become an important concern. In particular, verifying the identity of a satellite transmitter is now a basic requirement for protecting the services that rely on satellite access. In this article, we propose an active challenge-response authentication framework in which the verifier checks the satellite at randomly chosen times that are not known in advance, removing the fixed measurement window that existing passive methods expose to adversaries. The proposed framework uses the deterministic yet unpredictably sampled nature of orbital observables to establish a physics based root of trust for satellite identity authentication. This approach transforms satellite authentication from static feature matching into a spatiotemporal consistency verification problem inherently constrained by orbital dynamics, providing robust protection even against trajectory-aware spoofing attacks.
\end{abstract}

\section{Introduction}\label{sec1}

\subsection{Motivation}
In recent years, the rapid deployment of low Earth orbit (LEO) satellite constellations has significantly reshaped the landscape of global wireless communications.
Mega-constellations such as SpaceX Starlink, OneWeb, and Amazon Kuiper are placing thousands of satellites into LEOs, extending broadband access to areas that were previously out of reach. The 3rd Generation Partnership Project (3GPP) has also included non-terrestrial networks (NTNs) in the 5G New Radio standard from Release~17 onward, creating a common framework for satellite-assisted connectivity \cite{TR38821}. As the number of LEO satellites continues to grow, satellite networks are expected to support an increasing range of critical services such as emergency communications, navigation, data collection, and military operations. However, the proliferation of satellites also raises new security concerns, including the potential for satellite-based eavesdropping, i.e., unauthorized interception of communication signals \cite{my22TIFS}. Conversely, since satellites may potentially operate as malicious transmitters, verifying the identity of satellite transmitters has become a fundamental requirement for protecting services that rely on satellite access.

Authenticating a LEO satellite transmitter is considerably more difficult than in terrestrial networks. In conventional wireless systems, cryptographic key-based protocols can reliably verify the identity of a transmitter. However, satellite communication environments introduce several unique security challenges. First, satellites operate in remote orbital environments where physical access is extremely limited, making post-incident intervention difficult once a system is compromised \cite{Li20Satellite}. Second, satellite transmissions are inherently wide-area broadcasts, which significantly expands the attack surface and allows adversaries to observe and potentially inject signals from distant locations \cite{Suhaimi24State}. Third, advances in onboard software-defined radio and signal processing technologies have made it increasingly feasible for adversaries to imitate expected signal characteristics. 
% \del{For these reasons, relying solely on upper-layer cryptographic authentication may not be sufficient in satellite environments.} 
Therefore, these challenges have motivated growing interest in physical layer authentication (PLA), which leverages physical signal properties as an additional mechanism to verify transmitter identity.

% \sout{A more fundamental issue is that satellite signals are broadcast openly, which allows an adversary to transmit a forged signal that mimics a legitimate satellite transmission without holding any cryptographic key. These limitations of upper-layer security have driven growing interest in physical layer authentication (PLA), which uses the physical properties of the satellite channel, such as how propagation characteristics change over time, as an additional way to verify transmitter identity.}

% --- LaTeX Table Code Start (Refined for IEEE WCM) ---
\newcolumntype{M}[1]{>{\centering\arraybackslash}m{#1}}

\begin{table*}[t]
\centering
\renewcommand{\arraystretch}{1.5}

\begin{adjustbox}{max width=\textwidth}
\begin{tabular}{M{1.2cm}|M{3.2cm}|M{2.7cm}|M{4.2cm}|M{5.4cm}}
\specialrule{1.1pt}{0pt}{0pt}

% \rowcolor{mygreen!60}
\textbf{Reference} 
& \textbf{Physical Feature} 
& \textbf{Observation Window} 
& \textbf{Methodology} 
& \textbf{Limitation (Research Gap)} \\
\specialrule{1.1pt}{0pt}{0pt}

% ================= Data rows =================
% [Khan et al., 2024] \cite{khan2025modified} & Access Patterns + AKMA Key & Multiple / Passive & \textbf{Decentralized AKMA}: Validating IoT transmission patterns integrated with local security keys. & A hybrid approach dependent on upper-layer protocols; fails to exploit inherent orbital mechanics as a challenge. \\
% \hline
\cite{abdelaziz2019enhanced} & AoA & Single / Fixed & Spatial arrival angle authentication via antenna array estimation. & Relying on precise antenna hardware, rendering it susceptible to geometry-aware location spoofing. \\
\hline

\cite{fu2020initial} & Doppler frequency shift & Single / Fixed & Doppler based correlation using ephemeris derived reference values. & Dependent on single timestamp, failing to detect predictive signal mimicry and replay attacks. \\
\hline

\cite{topal2022physical} & Doppler frequency shift & Multiple / Fixed & Statistical Doppler spectrum fusion within inter-satellite links. & Lacking active authentication mechanisms, resulting in a failure to counter pre-calculated trajectory mimicking. \\
\hline

\cite{jedermann2021orbit} & TDoA signature & Multiple / Fixed & Matching arrival time patterns across distributed receivers. & Requiring a network of synchronized receivers and allowing predictive spoofing for fixed observation window. \\
\hline

\cite{oligeri2023past} & RF fingerprint & Multiple / Fixed & RF fingerprinting of hardware impairments through signal learning. & Sensitive to hardware aging, necessitating periodic model retraining and high computational overhead. \\
\hline

\cite{kumar2024authentication} & Atmospheric signature & Multiple / Fixed & ML-based profiling of atmospheric signal fluctuations. & Constrained by meteorological variability, leading to inconsistent performance across varying environments. \\
\hline

\cite{abdrabou2022physical} & Doppler + RSP  & Multiple / Sliding & SVM-based classification of Doppler and Received Power. & Ignoring Keplerian coupling between features and highly sensitive to environmental fading. \\
\hline

\rowcolor{yellow!20}
\textbf{Proposed Framework} & \textbf{Multi-Feature (Doppler, AoA, RSP, RTT)} & \textbf{Multiple / Random}  & \textbf{Active challenge-response verifying multi-feature orbital consistency.} & \textbf{Defeating predictive spoofing via randomized windows and enforcing strict physical feature coupling.} \\
\specialrule{1.1pt}{0pt}{0pt}

\end{tabular}
\end{adjustbox}

\vspace{0.4\baselineskip}
\caption{Summary of research on satellite physical layer authentication. (AoA: Angle of Arrival, TDoA: Time Difference of Arrival, RF: Radio Frequency, ML: Machine Learning, RSP: Received Signal Power, SVM: Support Vector Machine)}
\label{table_pla_comparison}
\end{table*}
% --- LaTeX Table Code End ---

\subsection{Taxonomy and Evolution of Satellite PLA}

% Existing satellite PLA methods have steadily improved over time, moving from simple single-measurement checks to more complex approaches that combine multiple signal features, as summarized in Table~\ref{table_pla_comparison}. Early work compared a single snapshot of the angle of arrival (AoA) \cite{abdelaziz2019enhanced} or Doppler frequency shift \cite{fu2020initial} against expected values computed from orbital data to check whether the signal matched the satellite's known position. Later works used multiple measurements at different times, such as Doppler statistics collected over inter-satellite links \cite{topal2022physical} or time difference of arrival (TDoA) matched across several ground stations \cite{jedermann2021orbit}, to track how the signal changes as the satellite moves. More recent studies have looked at hardware-level signal characteristics extracted from raw I/Q samples \cite{oligeri2023past} or signal fluctuations caused by the atmosphere \cite{kumar2024authentication}. Some approaches combine Doppler shift and signal strength together to make the check more reliable \cite{abdrabou2022physical}.

% To facilitate a comprehensive understanding of the current security landscape, existing satellite PLA methodologies can be categorized into four evolutionary stages based on their feature sets and observation windows, as summarized in Table~\ref{table_pla_comparison}. 

Existing satellite PLA techniques have evolved from simple single-feature checks toward more advanced schemes that exploit multiple signal features and temporal observations. To provide a structured view of this progression, existing methods can be categorized into four evolutionary stages based on their feature sets and observation windows, as summarized in Table~\ref{table_pla_comparison}.

\subsubsection{Single-Feature Authentication}
To provide low-overhead identity verification, early PLA schemes relied on the comparison of a single physical feature at a specific time instance. By estimating the angle of arrival (AoA) via antenna arrays \cite{abdelaziz2019enhanced} or correlating the Doppler frequency shift during initial access with ephemeris data \cite{fu2020initial}, these methods establish a basic root of trust. 
% \del{However, static schemes rely on a single snapshot and are vulnerable to collinear positioning attacks.} 
In such attacks, an adversary adjusts its spatial geometry to mimic the legitimate satellite, making one-time verification insufficient to detect sophisticated impersonation.

\subsubsection{Collaborative and Trajectory-Aware Authentication}
To address the limitations of single-snapshot verification, later studies introduced multi-point and sequence-based authentication. The time difference of arrival (TDoA) signatures are matched across synchronized ground receivers \cite{jedermann2021orbit}, while Doppler statistics are aggregated within inter-satellite links \cite{topal2022physical}. These methods incorporate temporal evolution into the authentication process. However, since the observation windows are deterministic and predictable, an adversary may be able to forecast the expected features and emulate the legitimate trajectory, which limits long-term security.

\subsubsection{Hardware and Environmental Fingerprinting}
Another research direction focuses on exploiting intrinsic device and channel characteristics. For example, hardware-specific impairments are extracted from I/Q samples \cite{oligeri2023past}, and spatiotemporal signal fluctuations caused by atmospheric propagation~\cite{kumar2024authentication} are analyzed. These methods rely on stochastic physical features for authentication. However, such signatures are sensitive to hardware aging and meteorological variations, which can lead to inconsistent performance across different operational environments.

\subsubsection{Multi-Feature Fusion and Physical Disconnect}
More recently, hybrid approaches have combined multiple physical features to improve authentication reliability in dynamic LEO environments. For example, Doppler shifts and received signal power (RSP) can be jointly analyzed to construct a richer identity profile \cite{abdrabou2022physical}. By fusing heterogeneous features, these schemes attempt to improve robustness against measurement noise and channel variations. However, the features are typically treated as independent statistical measurements rather than quantities determined by orbital dynamics. As a result, the physical relationships among signal features are not explicitly enforced during authentication. Consequently, even multi-feature fusion cannot fully eliminate spoofing opportunities when the observation timing is predictable.

\subsection{Contributions}

Despite these improvements, all existing approaches share a common weakness: they all rely on measuring signal features during a fixed, predictable time window. Because the timing of these measurements can be worked out in advance from publicly available orbital data, an adversary can may predict and send a carefully timed fake signal that matches these expected values throughout the measurement period. This weakness does not go away when more features are combined, because an adversary can align all of them at once if the timing is known ahead of time. Moreover, a deeper problem is that existing methods treat each measured feature as if it were independent from the others. In reality, under the laws of orbital motion, the Doppler shift, AoA, and round-trip signal delay of a satellite are all tied together; they are determined by the same orbital state at every moment. By ignoring this physical link, existing methods remain open to attacks in which an adversary copies the expected signal profile without actually being in orbit.

The key idea behind this work is that a LEO satellite's flight path, defined by six orbital parameters known as Keplerian elements, provides a built-in proof of identity that is enforced by the laws of physics. The way the signal features, specifically Doppler shift, AoA, and round-trip time (RTT), change over time is directly tied to the orbital path, so no transmitter can reproduce this pattern without actually flying on the same orbit at the same time. 
% \del{This physical constraint is inherent and cannot be easily bypassed through signal processing or by exploiting publicly available orbital data.}
Based on this idea, we propose an active challenge-response authentication framework in which the verifier checks the satellite at randomly chosen times that are not known in advance, removing the fixed measurement window that existing passive methods expose to adversaries. The satellite must respond consistently across all of these random check times, and any mismatch in the physical signal patterns reveals an impersonation attempt.

The main contributions of this work are as follows. First, we show that a satellite's orbital path is unique and use this uniqueness as a physical basis for authentication, clearly describing the physical link among Doppler shift, AoA, RSP, and RTT that makes this possible. Second, we design an active challenge-response protocol that replaces fixed-schedule passive measurement with randomly timed checks controlled by the verifier, blocking pre-planned impersonation attacks. Third, we build a multi-feature consistency check that verifies the physical coupling among all observed signal features at once, making it physically impossible to fake the correct response without occupying the same orbit. 
% The rest of this article is organized as follows. Section~II explains the orbital mechanics of LEO satellites and shows why Keplerian trajectories are unique and useful for security. Section~III describes the proposed active authentication framework. Section~IV presents a case study evaluating authentication performance, and Section~V discusses open research challenges and future directions for LEO satellite physical layer authentication.

\section{LEO Orbital Dynamics and Uniqueness}

\subsection{Keplerian Parameterization of LEO Orbits}

A LEO satellite's trajectory is fully and uniquely determined by six Keplerian orbital elements and the gravitational dynamics of Earth. In a geocentric inertial reference frame, the semi-major axis $a$ and eccentricity $e$ specify the size and shape of the orbital ellipse; the inclination $i$ defines the tilt of the orbital plane relative to the equatorial plane; the right ascension of the ascending node $\Omega$ fixes the orientation of that plane in inertial space; the argument of perigee $\omega$ orients the ellipse within the plane; and the true anomaly $\nu$ at a reference epoch locates the satellite on its orbit. 

\begin{figure}[!t]
\centering
\includegraphics[width=.75\columnwidth]{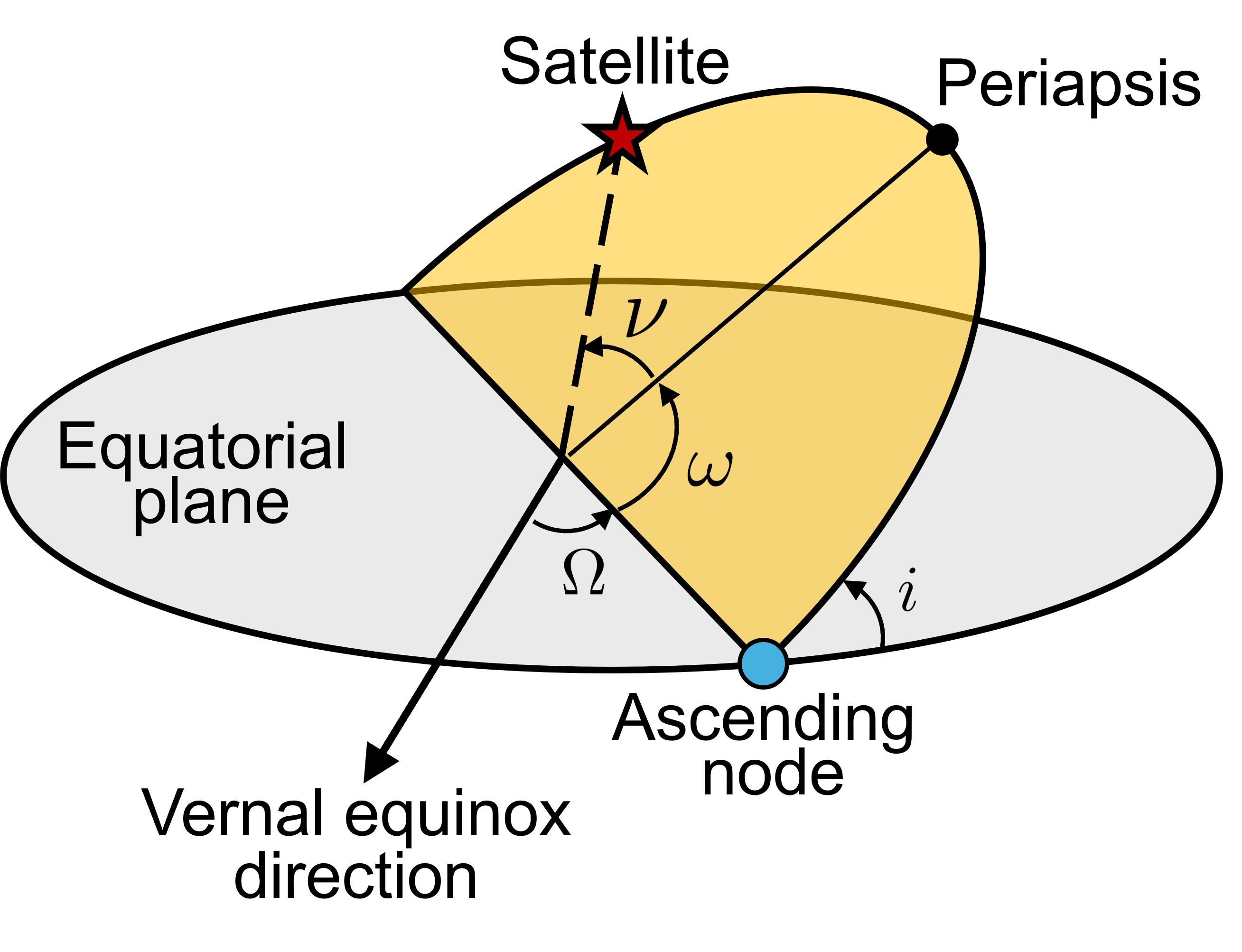}
\caption{Geometric illustration of the Keplerian orbital elements in the Earth-centered inertial (ECI) reference frame.}
\label{fig:orbital_elements}
\end{figure}

The geometric relationships among these six elements are illustrated in Fig.~\ref{fig:orbital_elements}, and the resulting six-tuple $\mathbf{o} = (a,\, e,\, i,\, \Omega,\, \omega,\, \nu)$ constitutes a complete and minimal state representation from which the satellite's position and velocity vectors can be propagated forward in time without ambiguity, using standard ephemeris models such as the Simplified General Perturbations 4 propagator applied to publicly available two-line element sets.
For the LEO altitude band of interest, orbits are typically near-circular with eccentricity $e \approx 0$. The orbital speed for a circular orbit can be obtained using the \textit{vis-viva equation}, i.e., $v = \sqrt{\mu/a}$, where $\mu = 3.986\times10^{14}~\mathrm{m^3/s^{2}}$ is the Earth's standard gravitational parameter. 
% \del{At an altitude of 550~km, the satellite's orbital speed is $v \approx 7.58~\mathrm{km/s}$, and the corresponding orbital period $T = 2\pi\sqrt{a^3/\mu} \approx 95.5~\mathrm{min}$. These} 
The high velocity of LEO satellites produces rapidly evolving channel conditions at any fixed ground receiver. 
% The precise temporal structure of these conditions is entirely encoded in $\mathbf{o}$ and is therefore predictable from publicly available ephemeris data.

% In practice, the Keplerian two-body model is augmented by perturbative forces that cause slow, predictable evolution of the orbital elements over time. The dominant perturbation in LEOs is the $J_2$ zonal harmonic of the Earth's geopotential, which arises from the planet's equatorial oblateness and induces secular regression of $\Omega$ and precession of $\omega$ at rates that depend sensitively on $a$, $e$, and $i$ \cite{Bevilacqua2008}. Atmospheric drag additionally causes a secular decay of $a$ at a rate determined by the satellite's ballistic coefficient, which is an intrinsic physical property that differs across spacecraft designs and cannot be replicated by a ground-based transmitter. Although these perturbations are slow on the timescale of a single pass, they continuously differentiate orbits that may share similar initial conditions, further reinforcing the long-term individuality of each satellite's trajectory.

\begin{figure*}[!t]
\centering
\includegraphics[width=\textwidth]{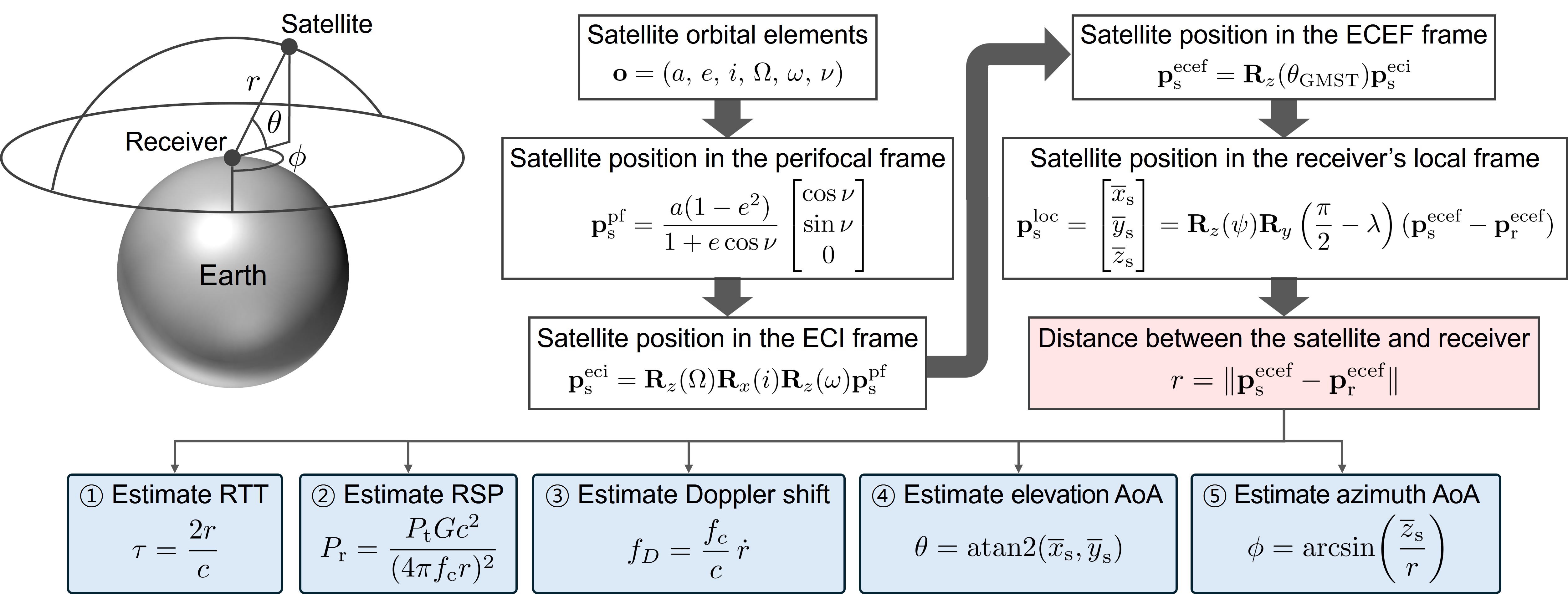}
\caption{Coordinate transformation chain from the perifocal orbital frame to the receiver's local frame and the resulting parameter estimations. The satellite position in the perifocal frame $\mathbf{p}_\text{s}^{\mathrm{pf}}$, parameterized by the true anomaly $\nu$, is first converted to the ECI frame through three intrinsic rotations about the $z$-$x'$-$z''$ axes by $\Omega$, $i$, and $\omega$, respectively, using elementary rotation matrices $\mathbf{R}_\alpha(\cdot)$, where $\alpha \in \{x, y, z\}$ denotes the axis of rotation. The ECI position is then transformed to the ECEF frame by rotating by the Greenwich mean sidereal time (GMST) angle $\theta_{\mathrm{GMST}}$, and subsequently projected into the local receiver frame defined by latitude $\lambda$ and longitude $\psi$. The slant range $r(t)$ between the satellite and the receiver is computed from the ECEF positions of the satellite $\mathbf{p}_\text{s}^{\mathrm{ecef}}$ and the receiver $\mathbf{p}_\text{r}^{\mathrm{ecef}}$. From $r$ and its time derivative $\dot{r}$, five observable features are estimated: RTT $\tau$, RSP $P_\text{r}$, Doppler shift $f_\text{D}$, elevation AoA $\theta$, and azimuth AoA $\phi$.}
\label{fig:coupling}
\end{figure*}

\subsection{Kinematic Observables and Their Coupling}

The relative kinematics between the satellite and a terrestrial receiver imprint four physically observable features onto the received signal: Doppler frequency shift $f_\text{D}(t)$, AoA decomposed into azimuth $\phi(t)$ and elevation components $\theta(t)$, RSP $P_\text{r}(t)$, and RTT $\tau(t)$. A defining property of these observables, and the one that underpins the security argument of this work, is that they are not statistically independent features. All four are deterministic functions of the same slant range, i.e., the distance between the satellite and receiver $r(t)$, and its time derivative $\dot{r}(t)$. These quantities are uniquely determined by~$\mathbf{o}$ via the coordinate transformation chain illustrated in Fig.~\ref{fig:coupling} and the ephemeris propagation equations.

The instantaneous Doppler shift for a carrier of frequency $f_\text{c}$ is $f_\text{D}(t) = (f_\text{c}/c)\,\dot{r}(t)$, where $c$ is the speed of light. Over the course of a pass, $f_\text{D}(t)$ traces a characteristic S-curve: large and positive as the satellite approaches, crossing zero at the point of closest approach, and large and negative as it recedes. The steepness of the transition, the epoch of the zero crossing, and the asymptotic amplitudes are uniquely determined by $\mathbf{o}$ and the receiver's geodetic coordinates. 
% At 550~km altitude, the peak Doppler excursion reaches approximately $\pm\,48~\mathrm{kHz}$ for an L-band carrier at 1.5~GHz. 
The AoA is obtained by transforming the satellite's Earth-centered Earth-fixed (ECEF) position vector into the receiver's local topocentric horizon frame: the elevation angle rises from near zero at the horizon, peaks at closest approach, and declines symmetrically, with its maximum value and timing uniquely prescribed by $a$ and the orbital geometry. The RSP is determined by the slant range through the Friis transmission equation as $P_\text{r} = P_\text{t} G c^2 / (4\pi f_\text{c} r)^2$, where $P_\text{t}$ is the transmit power and $G$ is the antenna gain. Since $r(t)$ varies continuously along the orbital path, the RSP follows a predictable temporal profile that peaks at closest approach and decays as the satellite moves toward the horizon. The RTT evolves as $\tau(t) = 2r(t)/c$, with time derivative $\dot{\tau}(t) = 2\dot{r}(t)/c$ that is directly proportional to the radial velocity. This last identity reveals the structural coupling: $f_\text{D}(t)$ and $\dot{\tau}(t)$ share the same physical origin $\dot{r}(t)$, so that any modification to one inevitably perturbs the other through a fixed algebraic relationship.

This inextricable coupling is the critical property that distinguishes satellite PLA from terrestrial counterparts. An adversary that attempts to forge the Doppler profile of a legitimate satellite must match $\dot{r}(t)$, which simultaneously fixes $\dot{\tau}(t)$ and constrains the transmitter's radial velocity relative to the receiver. 
% \del{This velocity constraint places the transmitter's required position and velocity on a locus that is geometrically identical to the legitimate orbital trajectory.} 
Satisfying all four constraints simultaneously is therefore not a signal processing problem but a physical placement problem: it requires the transmitter to reproduce the same kinematic trajectory.

\subsection{Trajectory Uniqueness as a Physical Security Primitive}

The foregoing analysis establishes that the six Keplerian elements $\mathbf{o}$ constitute a physics-enforced identity credential. Two satellites in distinct orbits cannot produce identical joint time series of $f_\text{D}(t)$, $\theta(t)$, $\phi(t)$, $P_\text{r}(t)$, and $\tau(t)$ at the same ground receiver. This is because the system of equations relating these observables to $\mathbf{o}$ uniquely determines the orbit given a sufficiently long observation window. Critically, this uniqueness is not probabilistic; it is a geometric consequence of the equations of motion. The orbital state vector therefore functions as a natural physical identifier that is imposed and continuously enforced by orbital mechanics rather than issued by a certificate authority.

This trajectory uniqueness becomes actionable as a security primitive through temporal observation. Since the orbital state at time $t$ is propagated deterministically from $\mathbf{o}$, the full pass-long trajectory $\{f_\text{D}(t),\, \theta(t), \, \phi(t),\, P_\text{r}(t),\, \tau(t)\}$ is fixed once $\mathbf{o}$ is specified. A verifier that compares observed features against an ephemeris-derived reference at a sequence of timestamps $t_1, t_2, \ldots, t_N$ accumulates kinematic evidence that grows in discriminative power with $N$. A spoofing transmitter may match the expected feature values at one timestamp, whether by chance or by pre-computation. However, if its physical trajectory deviates from the legitimate orbital path, unavoidable inconsistencies will appear at other timestamps because the four observables are coupled: adjusting one requires a physical repositioning that disturbs all others. Crucially, if the verification timestamps are chosen unpredictably by the verifier rather than disclosed in advance, the adversary cannot align its forged profile to the correct timestamps without real-time orbital placement. This observation motivates the core design choice of the authentication framework presented in the next section: an active challenge-response protocol in which the verifier selects randomized interrogation epochs, leveraging the deterministic yet unpredictably sampled nature of orbital observables as the physical root of trust for satellite identity authentication.

\begin{figure*}[!t]
\centering
\includegraphics[width=\textwidth]{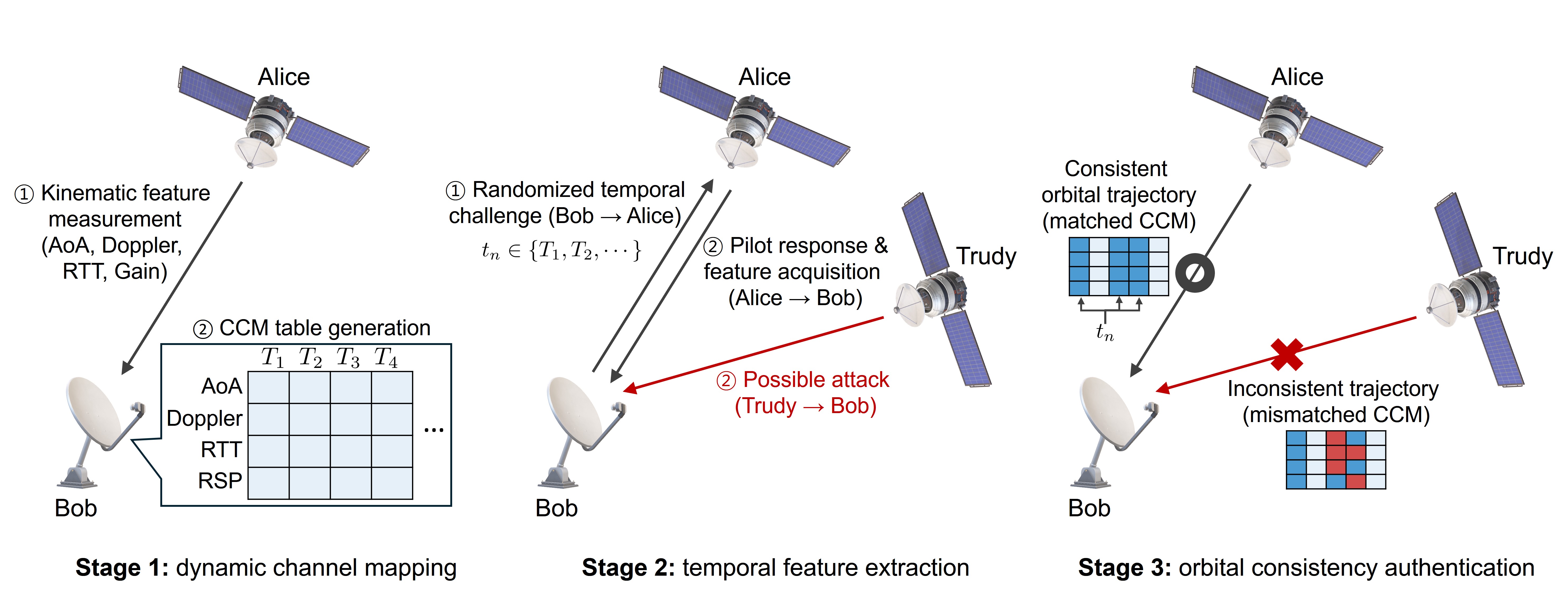}
\caption{Overview of the proposed active challenge-response authentication framework. Here, $T_1, T_2, \dots$ denote the absolute time slots within the satellite visibility window, and $t_1, t_2, \dots, t_N$ denote the $N$ time slots randomly selected from them and arranged in chronological order. In Stage~1, Bob measures kinematic features and constructs a CCM from satellite ephemeris data. In Stage~2, Bob issues randomized temporal challenges at $t_1, t_2, \dots, t_N$ and acquires multi-feature responses. An adversary, Trudy, may attempt to inject a forged signal. In Stage~3, Bob compares the observed features against the CCM: Alice exhibits a consistent orbital trajectory, while Trudy's response reveals kinematic inconsistencies that expose the impersonation attempt.}
\label{fig:proposed_cra}
\end{figure*}

\section{Proposed Framework: Active Spatiotemporal Authentication Exploiting Kinematic Uniqueness}
\label{sec:proposed_framework}

The authentication framework proposed in this work introduces a paradigm shift in securing LEO satellite networks, transitioning the defender's role from a conventional passive observer to that of an active verifier. Unlike existing PLA schemes that rely on the opportunistic measurement of channel features, the proposed framework exploits the deterministic nature of orbital mechanics to establish a physics based root of trust.

\subsection{Proposed Active Spatiotemporal Authentication}

The proposed framework performs authentication over multiple randomized timestamps, denoted as $t_1, t_2, \dots, t_N$. Since these time instances are selected unpredictably, authentication is no longer tied to a static spatial point but instead to a verifier-controlled dynamic trajectory. To realize this concept, the framework introduces an active spatiotemporal authentication mechanism based on a challenge-response protocol \cite{tomasin2022challenge, Piana25Challenge}. The verifier (Bob) determines the timing of each challenge, injecting temporal uncertainty into the process. This probing prevents adversaries from forecasting or synchronizing deceptive signal profiles in advance.

An adversary may replicate the satellite’s signal profile at a single, self-selected time instance. However, maintaining physical consistency across a randomized temporal sequence is significantly more challenging. The impersonator must simultaneously satisfy the coupled relationships among velocity, range, and geometry at every timestamp. If the adversary operates at a different altitude or follows a different velocity profile, it cannot satisfy Keplerian constraints. As a result, kinematic inconsistencies accumulate over time, exposing the impersonation attempt. Authentication can be strengthened when Bob moves, since randomized timestamps correspond to varying receiver positions. 
These variations are reflected in the authentication features, posing an additional challenge to Trudy, who is typically unaware of Bob's position.
However, Bob must know its position to compute the expected features.

To eliminate signal emulation, the proposed framework explicitly incorporates Keplerian orbital constraints into the authentication process. Specifically, Bob exploits the intrinsic coupling of satellite features through two hierarchical consistency checks, thereby establishing security that is rooted in orbital mechanics rather than feature-level statistics.

\subsubsection{Trajectory Uniqueness via Randomized Timestamps}
The proposed framework exploits the temporal evolution of satellite signals over multiple randomized timestamps to establish a reliable authentication basis. The verifier randomly selects the timestamps and evaluates the consistency of the resulting signal gradients across time. 
% \del{This process enables the system to distinguish legitimate orbital motion from spoofed signals.} 
As a result, matching features at a single timestamp is no longer sufficient for successful authentication. Under Keplerian dynamics, a satellite in a given LEO follows a unique kinematic trajectory. For example, it produces a characteristic Doppler S-curve and a specific rate of change in propagation delay. These temporal patterns are deterministically linked to orbital motion. By checking the consistency of signal features across randomized timestamps, the framework detects transmitters that do not follow the expected orbital trajectory. As observations accumulate over time, the resulting kinematic inconsistencies reveal the impersonation attempt.

\subsubsection{Multi-Feature Phsical Consistency}
To resist trajectory-mimicking attacks, the framework checks multiple physical features together instead of relying on a single signal property. Because orbital motion links Doppler shift, geometry, and propagation delay, these features must update consistently over time. A legitimate satellite naturally satisfies this relationship, whereas an impersonator cannot maintain such consistency. If an adversary attempts to replicate one parameter, such as the Doppler profile, it must adjust its relative velocity or transmission frequency. However, this adjustment inevitably affects other features, including AoA and RTT. For example, a ground-based spoofer may mimic the Doppler trend of a satellite, but the signal will arrive from a difference direction and with a different delay. By jointly validating these coupled features, the proposed scheme detects inconsistencies within forged signals. As a result, any transmitter that does not follow the true orbital trajectory cannot maintain consistent multi-feature behavior, making successful spoofing physically difficult.

\subsection{Overall System Operation}
This subsection presents  how the proposed framework operates within an active authentication process. As illustrated in Fig.~\ref{fig:proposed_cra}, the operational procedure consists of three main stages: dynamic channel mapping, temporal feature extraction, and orbital consistency authentication.

\subsubsection{Dynamic Channel Mapping}
The verifier first constructs a channel characteristic map (CCM) to provide a reliable authentication baseline. It derives dynamic reference data from satellite ephemeris and incorporates key propagation effects, such as AoA, Doppler shift, RTT, and RSP. By modeling the temporal evolution of these physical features, the system forms a ground-truth trajectory. This trajectory serves as a deterministic reference for legitimate signal paths throughout the satellite visibility window.

\subsubsection{Temporal Feature Extraction}
The active authentication process begins when Bob transmits a challenge that specifies randomized timestamps. Because the verifier controls the timing of each authentication instance, the process introduces temporal unpredictability and prevents adversarial pre-calculation. In response, the satellite (Alice) transmits pilot signals at the requested times. Then, the receiver samples the received signals and extracts a multi-feature vector that captures the instantaneous kinematic state of Alice.

\subsubsection{Orbital Consistency Authentication}
Bob makes the authentication decision by checking orbital consistency. It compares the observed feature sequence with the reference trajectory provided by the CCM. The similarity between them is measured using a metric that measures the cumulative difference between the observed features and reference trajectory. Impersonation attempts are detected by combining randomized temporal authentication with physics-based multi-feature consistency checks. 
% \del{Adversary that does not follow the expected orbital trajectory cannot simultaneously match coupled features such as Doppler shift, RTT, and AoA.} 
As these feature mismatches accumulate across multiple observations, the system exponentially reduces the detection error probability (DEP), effectively suppressing both false alarms and miss detection.
% \del{Compared with single timestamp-based authentication,} 
The proposed framework improves robustness by verifying consistency across multiple randomized observations. 
% \del{While an adversary may match the legitimate signal at a single timestamp. maintaining consistent behavior over time is physically difficult due to orbital constraints.} 
This highlights the key advantage of trajectory-based authentication, which relies on orbital motion rather than isolated feature matching.

\section{Case Study and Performance Analysis}

\begin{figure}[t]
\centering
\includegraphics[width=\columnwidth]{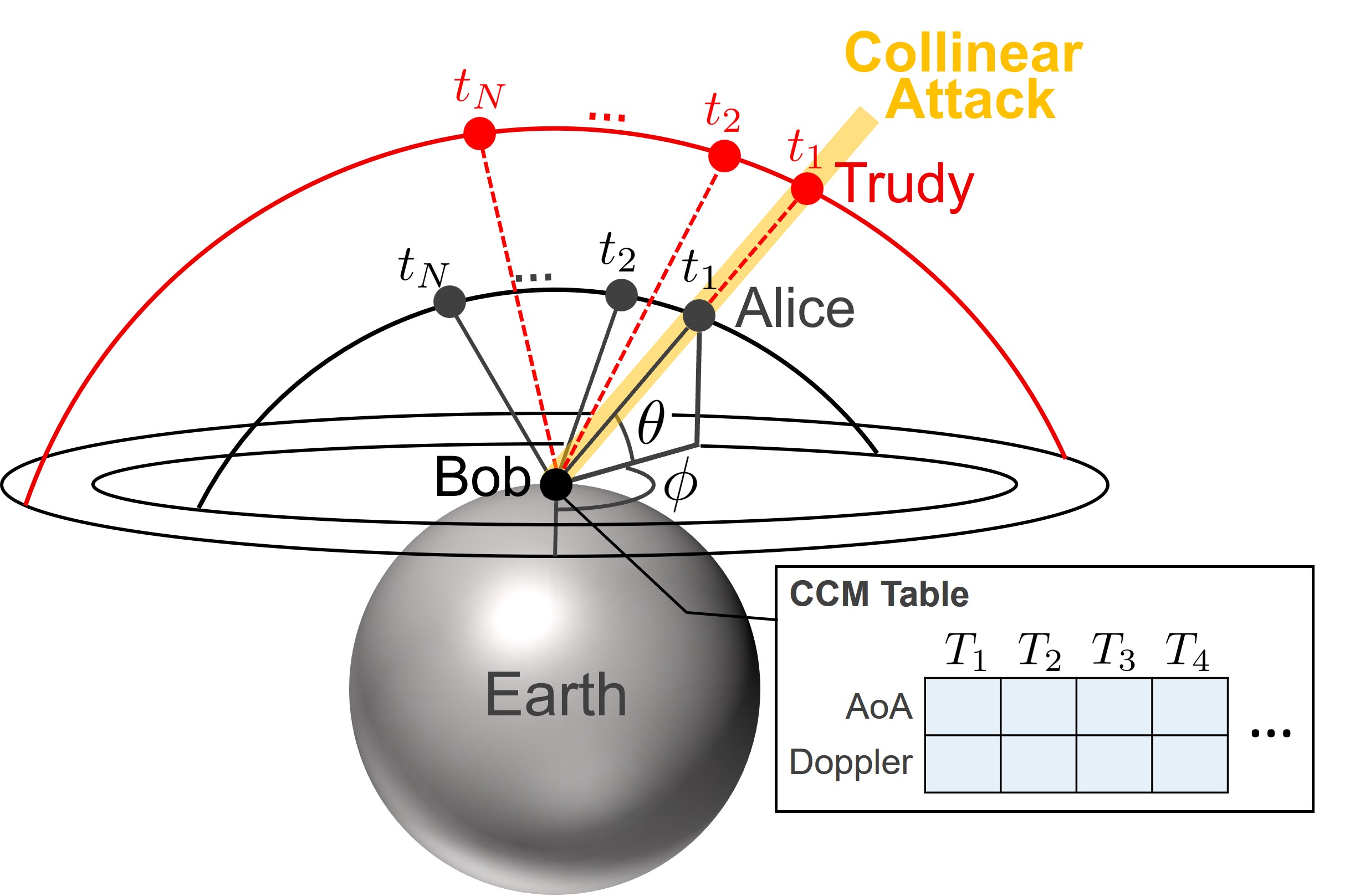}
\caption{Conceptual illustration of the collinear attack scenario and the verifier’s authentication model. Trudy attempts to match the AoA ($\theta$) of Alice at selected timestamp, e.g., $t_1,\dots,t_N$, while the verifier tracks the temporal consistency of physical features using the CCM table. In Scenario I (blind adversary), both AoA and Doppler are jointly used for multi-feature, multi-timestamp authentication. In Scenario II (informed adversary), Doppler is pre-compensated, and authentication relies on multi-timestamp AoA authentication.}
\label{fig:CollinearAttack}
\end{figure}

We evaluate the proposed framework with a legitimate satellite (Alice) at 600~km and an adversary (Trudy) at 1,200~km. Authentication is performed by tracking the temporal consistency of two coupled physical features, AoA and Doppler shift, across $N$ authentication timestamps $t_1, t_2, \dots, t_N$. 
To consider worst-case settings, we adopt the collinear attack as the primary threat model, as seen in Fig. \ref{fig:CollinearAttack}, where Trudy attempts to align her signal with Alice's AoA, i.e., $\phi$ and $\theta$, regardless of her level of orbital knowledge. We also consider two attack scenarios depending on the adversary's knowledge: a blind adversary without orbital information (Scenario~I) and an informed adversary with access to the satellite's ephemeris data (Scenario~II). The resulting authentication performance is measured by the minimum DEP, demonstrating the significant fusion gain achieved through multi-feature and multi-timestamp authentication. 

\begin{figure*}[!t]
\centering
\includegraphics[width=\linewidth, trim={1cm 0cm 1cm 0cm}, clip]{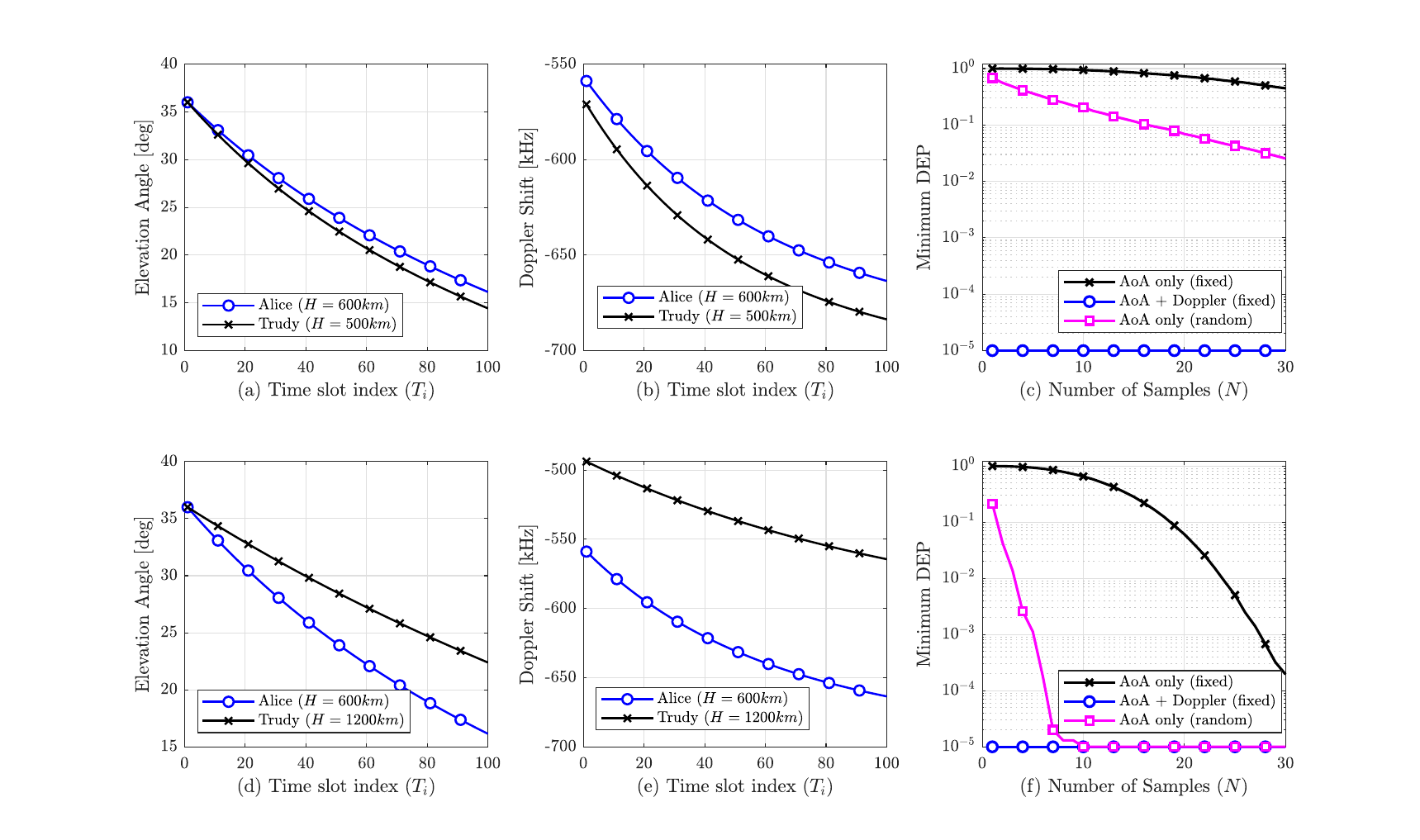}
\caption{Spatiotemporal trajectory analysis and authentication performance under different Trudy's altitudes (Top: $500$ km, Bottom: $1200$ km). (a) Elevation AoA trajectory mismatch, (b) Kinematic Doppler trajectory mismatch, and (c) Authentication performance (Minimum DEP) versus the number of observation samples $N$. Scenario I follows the ``AoA + Doppler (fixed)" curve, while Scenario II is restricted to the ``AoA only (fixed)" curve due to the Trudy's Doppler pre-compensation. Notably, Scenario III unpredictably selects $N$ timestamps across the entire time slot, following ``AoA only (random)" curve. This temporal randomness intrinsically captures the accumulated kinematic drift, drastically accelerating impersonator detection compared to conventional fixed sampling even when the initial spatial parameters are perfectly matched. For the performance evaluation, the measurement noise standard deviations of the elevation AoA and Doppler shift are set to $\sigma_{\theta} = 1.0 \text{ deg}$ and $\sigma_{f_D} = 200 \text{ Hz}$, respectively.}
\label{fig:AoA_Doppler_Perf}
\end{figure*}

\subsection{Scenario I: Blind Adversary with Fixed Timestamps}

We first consider a blind adversary that does not know Alice’s precise orbital ephemeris. Without the ability to synchronize her kinematic state, Trudy cannot replicate Alice's orbital motion. 
% \sout{Consequently, the received signal exhibits deterministic inconsistencies in both spatial and kinematic domaints, enabling immediate detection even in a single observation.}
As shown in Fig.~\ref{fig:AoA_Doppler_Perf}(a), a noticeable mismatch appears in the elevation AoA trajectory due to the orbital altitude difference. At the same time, without knowledge of Alice’s orbital velocity, Trudy cannot reproduce the correct Doppler shift evolution, which leads to the kinematic discrepancy illustrated in Fig.~\ref{fig:AoA_Doppler_Perf}(b). On the contrary, Bob can jointly utilize AoA and Doppler shift for authentication. The corresponding performance is represented by the ``AoA + Doppler (fixed)" curve in Fig.~\ref{fig:AoA_Doppler_Perf}(c). By combining these complementary physical features, the system achieves near-perfect authentication performance even with a single observation, $N=1$, since mismatches appear simultaneously across multiple physical features.

\subsection{Scenario II: Informed Adversary with Fixed Timestamps}

We next consider a knowledgeable adversary that has access to Alice’s orbital ephemeris. Using this information, Trudy can perfectly pre-compensate the difference between her own Doppler shift and that of Alice \cite{my24JSAC}, which masks the kinematic mismatch shown in Fig. \ref{fig:AoA_Doppler_Perf}(b). In addition, Trudy can achieve a collinear alignment with Alice and the ground station at a specific time instant, $t_1$, making the AoAs, e.g., $\theta$ and $\phi$, indistinguishable from the legitimate satellite at the beginning of the authentication process.
Under single timestamp-based authentication, such alignment can lead to successful impersonation. However, in the proposed active framework, this advantage is limited to the first observation. As the verifier performs a sequence of observations across multiple timestamps, e.g., $t_2, \dots, t_N$, Trudy gradually drifts away from the collinear configuration because her orbital angular velocity is lower than that of Alice. 

% \del{Indeed, the AoA has already been shown to be a robust feature for the PLA in other contexts.} 
In particular, AoAs are difficult to forge \cite{Pham2023}, even when Trudy can perform sophisticated signal processing and has many antennas, since it mostly depends on channel propagation conditions.
Since Doppler has already been compensated in this informed scenario, authentication mainly relies on AoA consistency. The resulting performance follows the ``AoA only (fixed)" curve in Fig. \ref{fig:AoA_Doppler_Perf}(c). Although the initial alignment at $t_1$ leads to a high authentication error and thus a high minimum DEP, the accumulation of trajectory evidence across multiple timestamps rapidly reduces the error. This result confirms that even when one feature is intentionally manipulated, temporal consistency of orbital geometry remains a reliable basis for authentication due to the uniqueness of satellite orbits.

\subsection{Scenario III: Informed Adversary with Randomized Timestamps}
Finally, we consider the informed adversary under the proposed randomized authentication strategy. In Scenario II, the adversary was able to align with Alice at a specific timestamp and partially imitate the signal geometry at the beginning of the authentication process. Although the subsequent observations eventually reveal the trajectory mismatch, the deterministic observation window still allows the adversary to predict the timing of Bob's authentication measurements.
To eliminate this predictability, the proposed framework randomly selects $N$ timestamps from the entire observation window rather than using deterministic observations. 
% \del{While Trudy can pre-compensate for the Doppler shift using orbital knowledge, she cannot reproduce the AoA consistency required across randomly selected timestamps without physically following the same trajectory.}
This randomized timestamp selection fundamentally limits the adversary's ability to maintain geometric alignment with Alice across all authentication instances. Although Trudy may achieve temporary collinear alignment at a specific instant, maintaining the same spatial relationship across multiple randomly-selected timestamps becomes physically infeasible due to the orbital dynamics of the satellites.
The resulting performance is illustrated by the ``AoA only (random)" curve in Fig.~\ref{fig:AoA_Doppler_Perf}(c). Even under the informed adversary model, the authentication error rapidly approaches zero as the number of sampled timestamps increases. This result highlights an important physical insight: 
% \del{although an adversary may locally mimic the signal characteristics at a particular moment,}
maintaining consistency with orbital geometry at multiple unpredictably selected timestamps becomes fundamentally constrained by orbital dynamics.

By introducing randomized multi-timestamp authentication, the proposed framework therefore transforms the authentication task from matching a single geometric snapshot to validating trajectory consistency at randomly selected timestamps. Consequently, the proposed active spatiotemporal authentication framework provides a robust defense against trajectory-aware spoofing attacks.

\subsection{Extension: Spatiotemporal Multi-Feature Coupling}
While Scenarios I--III primarily demonstrate authentication based on trajectory consistency observed through AoA and Doppler, satellite signals inherently exhibit spatiotemporal coupling across multiple physical features. In addition to AoA and Doppler, other orbit-dependent features such as RTT and RSP are also determined by the same underlying orbital dynamics. For example, randomized authentication timing directly constrains RTT spoofing. Each authentication request contains a unique challenge, forcing the adversary to generate the response only after receiving the signal. For an adversary located at a higher altitude, this requirement introduces a strict causality constraint imposed by signal propagation delay, making predictive pre-transmission infeasible.

The key advantage of the proposed framework lies in the joint effect of temporal unpredictability and the inherent coupling among multiple physical signal features determined by orbital motion.
% \del{Randomized authentication timestamps prevent the adversary from predicting when authentication will occur, while orbital dynamics tightly couple features such as AoA, Doppler, and RTT.}
Consequently, even if an adversary attempts to manipulate one feature, inconsistencies inevitably appear in others across randomly-selected authentication instants. This spatiotemporal constraint enables robust authentication even against highly capable adversaries with orbital knowledge.

\section{Open Challenges and Future Directions}
This section discusses key open challenges and future research directions for applying trajectory-based physical layer authentication in high-dynamic 6G NTNs. While the proposed active spatiotemporal authentication provides a robust root of trust, its large-scale deployment entails several open challenges that arise from the adversarial nature of the satellite-terrestrial interface.

% \subsection{Robust Verification under Jamming Attacks}
% The proposed framework depends on the reliable transmission of the challenge signal. This creates a potential vulnerability to jamming attacks. Since satellite passes are predictable, an adversary may attempt to interfere with the challenge transmission and prevent the satellite from responding. When the challenge signal is disrupted, the verifier cannot obtain the required kinematic response, which interrupts authentication. Moreover, because the challenge-response exchange occurs over an open wireless link, even partial jamming that corrupts a subset of the randomized timestamps can degrade the accumulated kinematic evidence and weaken the authentication decision. Future research should therefore explore anti-jamming signal and robust system designs that maintain reliable authentication under interference. Spread-spectrum challenge signals and frequency-hopping schemes that exploit the wideband nature of the satellite downlink can be promising candidates for ensuring challenge delivery even in contested spectrum environments.

\subsection{Adaptive Modeling for Non-Keplerian Dynamics}
The current framework assumes Keplerian orbital motion as the reference model. In practice, LEO satellites experience additional effects such as atmospheric drag and solar radiation pressure, which gradually alter their trajectories. Over time, these effects may cause mismatch between the reference CCM and the actual satellite motion, leading to unnecessary authentication failures. This issue becomes particularly important during periods of high solar activity, when increased atmospheric density accelerates orbital decay and amplifies the deviation from the nominal Keplerian prediction. Future work should investigate adaptive trajectory modeling techniques that continuously refine the reference trajectory using updated ephemeris information. Incorporating real-time orbit determination data and state estimates obtained via onboard global navigation satellite systems into the CCM update process could significantly reduce reference model errors while preserving the physical consistency required for robust authentication.

% \tcr{\subsection{Scalability Toward Massive MIMO and Beam-Hopping Architectures}}
% Future LEO satellite systems are expected to adopt massive MIMO and dynamic beam-hopping technologies to support high-capacity connectivity. In these environments, authentication must operate across many directional beams at the same time. Trajectory-based authentication can naturally benefit from beamforming diversity since spatially separated beams provide additional physical information. However, performing randomized authentication over a large number of beams introduces considerable scheduling and coordination complexity. Therefore, future research should focus on lightweight and beam-aware authentication mechanisms that minimize overhead while maintaining reliable authentication performance. Such designs will be essential for scalable deployment in large-scale mega-constellation systems. This highlights the importance of scalable authentication designs that evolve together with future beam-centric NTN architectures.

\subsection{Verification Overhead and Latency in Delay-Sensitive Services}
The proposed framework requires multiple challenge-response exchanges across randomized timestamps to accumulate sufficient kinematic evidence. Increasing the number of verification instances $N$ improves authentication reliability, but each exchange consumes part of the limited satellite visibility window and adds communication latency. For delay-sensitive services such as emergency communications and navigation, this creates a fundamental tradeoff between authentication accuracy and service responsiveness. At a typical LEO altitude of 550 km, a satellite pass lasts only a few minutes, which limits the time available for both authentication and data transmission. Future work should explore adaptive verification strategies that adjust $N$ based on the required security level and the latency constraints of the target service. Simplified challenge designs that minimize round-trip overhead while preserving the temporal unpredictability of the framework will also be essential for practical deployment.

\subsection{Authentication Continuity Across Satellite Handovers}
In LEO mega-constellations, frequent satellite handovers are unavoidable because each satellite remains visible for only a few minutes. Whenever a handover occurs, the serving satellite changes, and the verifier must authenticate a new satellite. This repeated re-authentication introduces additional signaling overhead and temporarily leaves the link unverified during the transition period. 
The challenge becomes more severe in dense constellation scenarios where handovers occur more frequently, requiring the verifier to complete a full multi-timestamp verification cycle for each new satellite within a short contact window. 

Future research should therefore investigate mechanisms that maintain authentication continuity across handovers. For example, partial authentication evidence from a previous satellite could be reused to accelerate the verification process for the next satellite. Another promising direction is cooperative authentication, where neighboring satellites or ground stations share orbital consistency information with the verifier to reduce the authentication burden during rapid handover sequences. Such mechanisms could enable trajectory-based authentication to operate efficiently in large-scale LEO constellations.

\subsection{Effects of Bob's Movement}
Further investigation is needed on the effects of Bob's mobility on PLA performance. Considering contexts in which such mobility is possible, designing proper random movement strategies to increase the variation and unpredictability of the feature would make impersonation attacks more difficult. Additionally, an analysis of performance with each feature should be conducted, taking into account specific propagation conditions. This would reveal the conditions under which physical features such as the AoA and Doppler shift cannot be forged by Trudy.

\bibliographystyle{IEEEtran}
\bibliography{reference}

\section*{Biographies}
\begin{IEEEbiography}{Jinyoung Lee} (haetsal120@gmail.com) is an assistant professor at National Korea Maritime \& Ocean University, South Korea. He previously held a staff engineer at Samsung Electronics. His current research focuses on physical-layer security, including authentication and covert transmissions, with applications in UAVs, LEO satellites and distributed AI. 
\end{IEEEbiography}
\begin{IEEEbiography}{Stefano Tomasin}  (stefano.tomasin@unipd.it) is a full professor at the University of Padova, Italy. His interests include signal processing for communications and physical layer security. From 2020 to 2023, he was an Editor of the IEEE Transactions on Information Forensics and Security, and from 2023, he is deputy editor in chief of the same journal.
\end{IEEEbiography}
\begin{IEEEbiography}{Dong-Hyun Jung} (dhjung@ssu.ac.kr) is an assistant professor at Soongsil University, South Korea. From 2017 to 2025, he was a senior researcher at Electronics and Telecommunications Research Institute, South Korea. His research interests include satellite communications, satellite clustering, and physical-layer security in non-terrestrial networks.
\end{IEEEbiography}

\end{document}